\documentclass[aps,pre,twocolumn,showpacs]{revtex4}
\usepackage{epsfig}
\usepackage{times}
\begin{document}

\title{Cooperation in noisy case: prisoner's dilemma game on two types of regular random graphs}

\author{Jeromos Vukov$^1$, Gy\"orgy Szab\'o$^2$, and Attila Szolnoki$^2$}
\affiliation {$^1$Department of Biological Physics, E\"otv\"os
University, H-1117 Budapest, P\'azm\'any P. stny. 1/A.,
Hungary  \\
$^2$Research Institute for Technical Physics and Materials Science
P.O. Box 49, H-1525 Budapest, Hungary}

\begin{abstract}
We have studied an evolutionary prisoner's dilemma game with
players located on two types of random regular graphs with a
degree of 4. The analysis is focused on the effects of payoffs and
noise (temperature) on the maintenance of cooperation. When
varying the noise level and/or the highest payoff, the system
exhibits a second order phase transition from a mixed state of
cooperators and defectors to an absorbing state where only
defectors remain alive. For the random regular graph (and Bethe
lattice) the behavior of the system is similar to those found
previously on the square lattice with nearest neighbor
interactions, although the measure of cooperation is enhanced by
the absence of loops in the connectivity structure. For low noises
the optimal connectivity structure is built up from randomly
connected triangles.
\end{abstract}


\date{\today}

\maketitle

One of the central questions in evolutionary game theory is to find necessary conditions and mechanisms which result in cooperation among selfish individuals. Naturally, several mechanisms have already been explored such as kin selection \cite{hamilton_jtb64b}, retaliating behavior \cite{axelrod_84}, voluntary participation \cite{hauert_s02}
or development of reputation \cite{fehr_n03}.

A widely studied toy model, which illustrates the conflict between
cooperation and selfish behavior is the prisoner's dilemma game
(PDG). In the (two-player and one-shot) PDG \cite{weibull_95,gintis_00}, the players simultaneously make decision whether cooperate or defect to maximize their individual payoff. The dilemma is based on the fact that defection brings higher income independently of the other player's decision. But if both players defect they receive significantly lower payoff than in the case of mutual cooperation.

The introduction of short-range interaction between the spatially distributed players can explain the formation of cooperation for the iterated evolutionary PDGs \cite{nowak_ijbc93,nowak_ijbc94}, even for the case when players can follow one of the two simplest strategies as ``always cooperate'' ($C$) and ``always defect'' ($D$). In multi-agent evolutionary PDGs, the players gain their income from games with their neighbors. According to the Darwinian selection principle, the less successful strategies are replaced by more successful strategies adopted from their neighborhood \cite{maynard_82,hofbauer_98}.

Lots of studies followed the pioneering work of Nowak {\it et al.}
\cite{nowak_ijbc93,nowak_ijbc94} in the field of evolutionary PDGs
analyzing many types of connectivity structures (lattices
\cite{lindgren_pd94,nakamaru_jtb97,szabo_pre98}, diluted lattices \cite{nowak_ijbc94}, social networks \cite{abramson_pre01,ebel_pre02,kim_pre02,masuda_pla03},
hierarchical graphs \cite{vukov_pre05}, scale-free networks \cite{santos_prl05}, preferential selection of a neighbor \cite{wu_cpl06}) to clarify the possible role of the topology. In Ref.~\cite{zimmermann_pre04}, self organizing
networks controlled by proper dynamics were applied to find the
most advantageous structure for cooperation. Our previous study~\cite{szabo_pre05} reveals the importance of clique percolation~\cite{derenyi_prl05,palla_n05}.

In the present paper, as a continuation of our previous work
\cite{szabo_pre05}, we explore the effect of noise parameter on the PDG by extending the connectivity structures to random graphs. Our observations suggest that the effects of noise (temperature) on the stationary concentration of cooperators is greatly affected by the underlying structure.

To investigate the influence of the noise and connectivity structures, we have considered an evolutionary PDG with players located on the sites of a regular graph with a connectivity of four ($z=4$). The restriction to regular random graph serves to make the comparison easier and to avoid undesired effects due to the various degrees of nodes (different sizes of neighborhoods) \cite{nowak_ijbc94,schweitzer_acs02,duran_pd05,santos_prl05}.

The players can follow one of the above mentioned two strategies whose distribution is described by the formalism of the two-state Potts model: the possible state of the site $x$ is ${\bf s}_x=C$ or $D$ (cooperator or defector). The strategy adoption mechanism \cite{szabo_pre05} is based on the rescaled version of the payoff matrix introduced by Nowak {\it et al.} \cite{nowak_ijbc93}:
\begin{equation}
%\label{eq:pom}
{\bf A}=\left( \matrix{0 & b \cr
                       0 & 1 \cr} \right)\;, \;\; 1 < b < 2 \;.
\end{equation}
During the evolutionary process the randomly chosen player at site $x$ can adopt one of the (randomly chosen) co-player's (at site $y$) strategy with a probability depending on the payoff difference $(U_x-U_y)$
\begin{equation}
\label{eq:update} W[{\bf s}_x \leftarrow {\bf s}_y] = {1 \over 1 +
 \exp {[(U_x-U_y)/K]} } \;,
\end{equation}
where $K$ characterizes the magnitude of noise involving many different effects (fluctuations in payoffs, errors in decision, individual trials, etc.) \cite{blume_geb93,szabo_pre98}.

\begin{figure}[ht]
\centerline{\epsfig{file=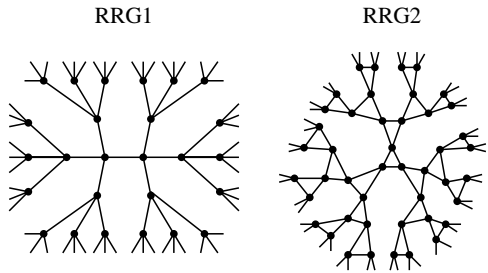,width=7cm}}
\caption{\label{fig:struc}Two types of random regular graphs on which an evolutionary Prisoner's Dilemma game is studied.}
\end{figure}
The present analysis is focused on two connectivity structures displayed in Fig.~\ref{fig:struc}. Henceforth, we will refer to the structures as RRG1 and RRG2, respectively. The RRG1 random regular graph is the simulated version of the Bethe-lattice, because RRG1 is locally similar to a tree in the large size limit (when the number of sites goes to infinity). For finite number $N$ of sites, however, RRG1 has loops. This influence seems to be negligible when choosing sufficiently large systems. To avoid confusion and for better visualization, Fig.~\ref{fig:struc} represents an ``ideal'' (free of loops) part of RRG1.

RRG2 is a random regular graph of triangles with three neighboring triangles. The triangles of RRG2 are the nodes of the underlying random regular graph (with $z=3$), and between these nodes, the bonds are the common sites of the overlapping triangles. Locally, it resembles to the Kagome lattice, but -- as will be seen later -- it has some characteristics of a random regular graph, too. Despite the similar coordination number ($z=4$) the topology of the two random networks differs significantly. In the limit $N \to \infty$, the concentration of triangles (three sites cliques) vanishes in RRG1, {\it i.e.}, this structure has a clustering coefficient ${\cal C}=0$, whereas the overlapping triangles percolate on RRG2, and the clustering coefficient
is ${\cal C}=1/3$.

Since the classical mean-field theory is insensible to the topology, the equation of motion for the concentration of cooperators is identical with the equation of the model on lattices presented in our previous work \cite{szabo_pre05}. The solution of differential equation suggests that cooperators die out and defection is the only successful strategy for arbitrary values of $K$ and $b>1$.

As it was mentioned in our previous paper \cite{szabo_pre05} the necessary condition for cooperators to survive is the possibility to form clusters where cooperators can assist each other. This may happen on lattices in higher dimensions at certain range of parameter $b$. The present paper surveys the areas of the $b$-$K$ parameter plain where cooperation can survive.

\begin{figure}[ht]
\centerline{\includegraphics[angle=-90,width=8cm]{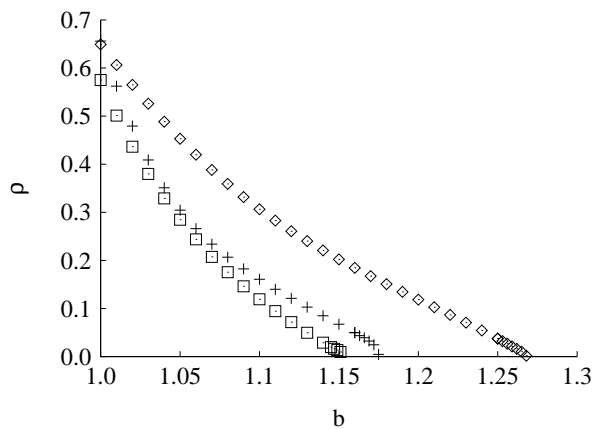}}
\caption{\label{fig:rrgmcs}Monte Carlo results for the concentration of cooperators {\it vs}. $b$ for three different noise values: $K=0.1$ (pluses), 0.3 (diamonds), and 1.2 (squares) on RRG1.}
\end{figure}

Figure~\ref{fig:rrgmcs} shows the stationary concentration $\rho$ of cooperators on RRG1 as a function of $b$ for different temperature values. This plot indicates that the variation of $\rho$ on RRG1 is qualitatively similar to those found on the square lattice~\cite{szabo_pre05}. These numerical data are obtained from Monte Carlo (MC) simulations performed on large systems [the size varied from $N=4 \cdot 10^4$ to $N=4 \cdot 10^6$], and the stationary values are determined by averaging over a sampling time $t_{s}$ [varied from $t_s=10^4$ to $10^6$ Monte Carlo steps per sites]. The larger $N$ and $t_s$ are used in the close vicinity of the extinction of cooperators because, the variance diverges approximately as $\chi \propto \rho^{-2}$ for small concentration and low $K$. Further difficulties were caused by the size dependence at low $K$ values. The reason of this phenomenon can be related to the existence of small loops whose effect will be discussed later on.

The transition from a fluctuating ($C+D$) phase into the absorbing state
frequently belongs to the directed percolation universality class.
The critical exponent characterizing the power law decrease of concentration were reproduced when the evolutionary PDG was simulated on different two-dimensional lattices \cite{szabo_pre98,szabo_pre05}. As expected, the change of host lattice to random graph results in a mean-field type behavior in $\rho(K)$ as demonstrated in Fig.~\ref{fig:rrgmcs}. The stationary concentration of cooperators is independent of the initial state and decreases linearly with $b_{cr}-b$. As $b$ exceeds a threshold value ($b_{cr}$), cooperation cannot be maintained, the evolution always ends in the homogeneous $D$ state.

\begin{figure}[ht]
\centerline{\includegraphics[angle=-90,width=8cm]{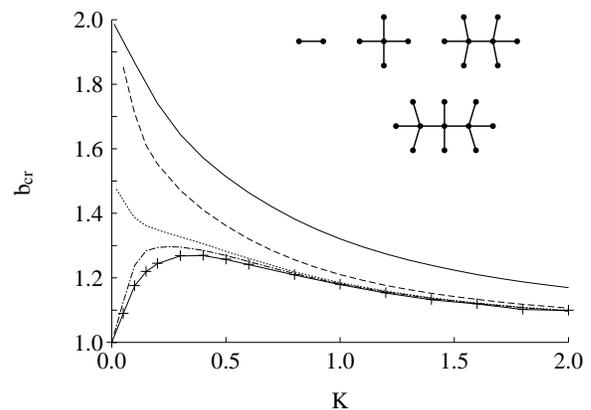}}
\caption{\label{fig:rrgtbc}Critical value of $b$ {\it vs}. $K$ on RRG1. Symbols come from Monte Carlo simulations,
the solid, dashed, dotted, and dashed-dotted lines represent the
predictions of dynamical cluster approximation for 2-, 5-, 8-, and
11-site clusters shown at the top.}
\end{figure}

We have determined the critical values of $b$ for many different noises and the results are illustrated in Fig.~\ref{fig:rrgtbc}. The $b_{cr}(K)$ curve has a maximum at $K \approx 0.37$, and goes to 1 if $K$ goes either to zero or to infinity. This figure can be considered as a phase diagram because the solid line connecting the Monte Carlo data separates the area where cooperators and defectors can coexist from the pure defector area.

The possibility of a similar stochastic resonance was reported by Traulsen {\it et al}.~\cite{traulsen_prl04} who considered another evolutionary rule based on the application of the "win-stay-lose-shift" strategies. The appearance of stochastic resonance was directly demonstrated on the square lattice by Perc in a similar PDG with adding random perturbations to the payoffs~\cite{perc_njp06}.

The analytical reproduction of this behavior proved to be a very
time-consuming task. In contrary to the Monte Carlo results, as
detailed in the previous paper \cite{szabo_pre05}, the traditional
mean-field approximation suggests a sudden change between the
homogeneous $D$ and $C$ states at $b_{cr}^{(mf)}(K)=1$. The pair
approximation is capable to describe the coexistence of the $C$
and $D$ strategies but it gives a rough estimation for critical
$b$ values, especially in the zero temperature limit. To eliminate
these discrepancies we have to extend the dynamical cluster
approximations (considered as generalized mean-field methods) on
the Bethe lattice~\cite{szabo_pre00b}. When applying these
techniques we derive equation of motion for all possible
configurational probabilities on large clusters and search
numerically for the stationary concentrations by integrating the
equations of motion with respect to time (further details on these
methods will be given elsewhere). Figure~\ref{fig:rrgtbc} shows
that both the five- and eight-site approximations predict
incorrect results in the zero temperature limit although their
predictions become more and more accurate in the high temperature
region. Besides it, one can observe relevant improvement when
comparing the results of the five- and eight-site approximations
at the limit $K \to 0$. This fact inspired us to extend this
method for the level of 11-site approximations. As shown in
Fig.~\ref{fig:rrgtbc} this level is already capable of describing
the disappearance of cooperation as $K$ goes to $0$. The above
series of results emphasizes the importance of long-range
correlations and/or the absence of loops in the connectivity
structures for the limit $K \to 0$.

\begin{figure}[ht]
\centerline{\includegraphics[angle=-90,width=9cm]{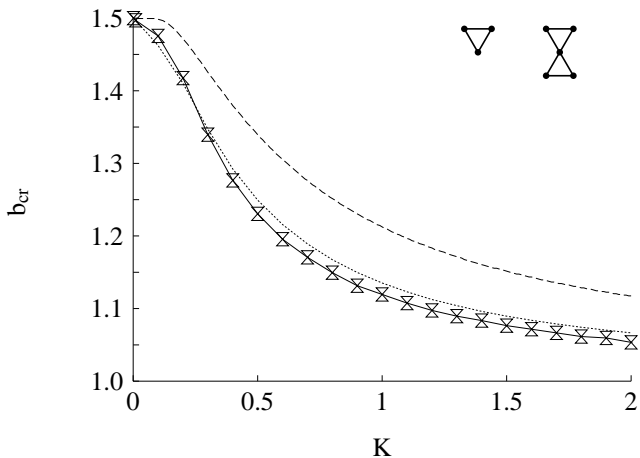}}
\caption{\label{fig:krrgtbc}Phase diagram on RRG2. Symbols denote
the MC data. The dashed and dotted lines illustrate the phase
boundary between the $D$ and $(C+D)$ phases predicted by the 3-
and 5-site approximation on the clusters shown at the top.}
\end{figure}

Basically different behavior is found on the RRG2 structure as demonstrated in Fig.~\ref{fig:krrgtbc}. It is conjectured previously that the function $b_{cr}(K)$ decreases monotonously to 1 if $K$ is increased for those connectivity structures where the overlapping triangles spans the whole system. The present data support this conjecture.

\begin{figure}[ht]
\centerline{\includegraphics[angle=-90,width=9cm]{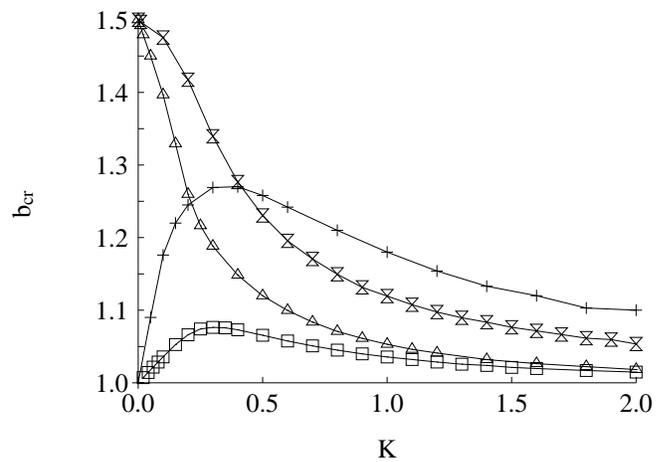}}
\caption{\label{fig:tbc}Phase boundaries between the $D$ and
$(C+D)$ phases on RRG1 (pluses), RRG2 (double-triangle symbols),
square lattice (squares), and Kagome lattice (triangles).}
\end{figure}

In Fig.~\ref{fig:krrgtbc} we compare the MC data with the
predictions of three- and five-site approximations. It is worth
mentioning that qualitatively similar results are obtained when
the connectivity structure is given by the Kagome lattice
\cite{szabo_pre05}. The RRG2 and Kagome lattice are locally
similar: the overlapping triangles have only one common site. This
is the main reason why the three- and five-site approximations
predict the same results on the RRG2 and Kagome lattice. In fact,
the structure of RRG2 fits well to the conditions of the five-site
approximations overestimating both $\rho$ and $b_{cr}$ for the
Kagome lattice. At the same time, the results of this approach
agrees very well with the MC data on RRG2 (see
Fig.~\ref{fig:krrgtbc}). We have to emphasize that this is the
connectivity structure providing the highest measure of
cooperation among regular structures (if $z=4$) for low $K$ values
(for a comparison see Fig.~\ref{fig:tbc}). At the first glance it
is a surprising result because the introduction of spatial
connectivity structures was motivated by the possibility of the
formation of $C$ colonies \cite{axelrod_84,nowak_ijbc93}.

The above results have helped us to deduce a simple explanation
justifying the importance of the one-site overlapping triangles in
the connectivity structures. Let us assume that one of the
triangles is occupied by cooperators in the sea of defectors.
Within this triangle, the cooperator's income is 2, the
neighboring defectors receive $b$, and all the other $D$s get
nothing. In this situation the most probable evolutionary process
is that one of the neighboring defectors adopts the strategy of
the more successful cooperators (for low $K$). The state of this
new cooperator is not stable and it can be switched into defector
again within a short time. During the lifetime of the new
cooperator, however, the other neighboring defector adopts the $C$
strategy very probably from the cooperator in the original
triangle (if $b < 3/2$), and they will form a neighboring (stable)
triplet of cooperators. The iteration of these processes yields a
growing tree of cooperator triplets. The growing process is
stopped at the sites which separate two branches of the growing
tree, because the corresponding defector(s) can exploit two or
more cooperators simultaneously. The blocking events occur
frequently for the spatial structures and are excluded for
tree-like structures. Thus, the absence of loops (formed by the
one-site overlapping triangles) sustain the spreading of
cooperation in RRG2. This picture is valid if only one tree of
cooperator triplets exists in the initial state. For random
initial state, however, the system will have many (tree of)
cooperator triplets which will check each other in the growing
process at the sites where their branches are separated even on
the RRG2 structure. In the final stationary state the spreading
and blocking effects are balanced in a way taking the noise into
account. Evidently, the noisy effects can break up the triplets of
cooperators. As a result, one can observe monotonous decrease in
the function $\rho(K)$ when increasing $K$ for any $b < b_{cr}$.

Disregarding the triangles, the presence of loops in the connectivity structure reduces the measure of cooperation for high noises too. Figure~\ref{fig:tbc} clearly demonstrates that RRG2 is more advantageous than the Kagome lattice in the maintenance of cooperation. Similar conclusion can be deduced when we compare the results obtained on the RRG1 and square lattice (see Fig.~\ref{fig:tbc}). In these cases the stable triplets of cooperators cannot be observed because of the absence of triangles in the connectivity structure. For both structures the maintenance of cooperation is supported by the noisy events resembling to the stochastic resonance \cite{perc_njp06}. For finite noises the occasional (irrational) adoption of the neighboring $C$ strategy increases the tree of cooperators, and this effect is weakened by the presence of loops as described above.

It is more surprising that the disadvantageous presence of loops (with length longer than 3) is so relevant that RRG1 becomes the most efficient structure to sustain the cooperation if $K$ exceeds a threshold value ($K > K_{th} \simeq 0.4$) for $z=4$. In other words, in the maintenance of cooperation the advantageous effect of the quenched (regular) neighborhood is weakened by the spatial characteristics of the connectivity structure.

To summarize, we have systematically studied the effect of
noise $K$ and temptation $b$ to choose defection on the measure of cooperation in an evolutionary PD game for two types of random regular connectivity structures with $z=4$. For this purpose we have determined the critical value $b_{cr}(K)$ of temptation until the cooperators can remain alive. For sufficiently high noise levels the comparison of different connectivity structures indicates that the highest $b_{cr}(K)$ can be achieved by minimizing the number of loops in the connectivity structure. On the contrary, at low noise levels, the preferred structure is built up randomly from overlapping triangles in a way that the overlapping triangles have only one common site. Evidently, the analysis becomes more complicated for those connectivity structures which involve variation in degree, $z>4$, and more complex connections at the overlapping triangles.

\begin{acknowledgments}

This work was supported by the Hungarian National Research Fund
(Grant No. T-47003) and by the European Science Foundation (COST P10).

\end{acknowledgments}

%\bibliography{egg}

\end{document}